\documentclass[]{spie}  

\usepackage{amsmath,amsfonts,amssymb}
\usepackage{graphicx}
\usepackage{multirow}
\usepackage[colorlinks=true, allcolors=blue]{hyperref}

\title{Monte Carlo simulations of soft proton flares: \\testing the physics with XMM-Newton}

\author[a]{Valentina Fioretti}
\author[a]{Andrea Bulgarelli}
\author[a]{Giuseppe Malaguti}
\author[b]{Daniele Spiga}
\author[c]{Andrea Tiengo}

\affil[a]{INAF Istituto di Astrofisica Spaziale e Fisica Cosmica Bologna, Via P. Gobetti 101, 40129, Bologna, Italy}
\affil[b]{INAF Osservatorio Astronomico di Brera, Via E. Bianchi 46, 23807 Merate (LC), Italy}
\affil[c]{INAF Istituto di Astrofisica Spaziale e Fisica Cosmica Milano, via Bassini 15, 20133 Milano, Italy}

\authorinfo{Further author information: (Send correspondence to V. Fioretti)\\E-mail: fioretti@iasfbo.inaf.it, Telephone: +39 051 6398772}

\begin{document} 
\maketitle

\begin{abstract}
Low energy protons ($<100-300$ keV) in the Van Allen belt and the outer regions can enter the field of view of X-ray focusing telescopes, interact with the Wolter-I optics, and reach the focal plane. 
The funneling of soft protons was discovered after the damaging of the Chandra/ACIS Front-Illuminated CCDs in September 1999 after the first passages through the radiation belt. The use of special filters protects the XMM-Newton focal plane below an altitude of 70000 km, but above this limit the effect of soft protons is still present in the form of sudden flares in the count rate of the EPIC instruments that can last from hundreds of seconds to hours and can hardly be disentangled from X-ray photons, causing the loss of large amounts of observing time.
The accurate characterization of (i) the distribution of the soft proton population, (ii) the physics interaction at play, and (iii) the effect on the focal plane, are mandatory to evaluate the background and design the proton magnetic diverter on board future X-ray focusing telescopes (e.g. ATHENA). 
Several solutions have been proposed so far for the primary population and the physics interaction, however the difficulty in precise angle and energy measurements in laboratory makes the smoking gun still unclear.
Since the only real data available is the XMM-Newton spectrum of soft proton flares in orbit, we try to characterize the input proton population and the physics interaction by simulating, using the BoGEMMS framework, the proton interaction with a simplified model of the X-ray mirror module and the focal plane, and comparing the result with a real observation.
The analysis of ten orbits of observations of the EPIC/pn instrument show that the detection of flares in regions far outside the radiation belt is largely influenced by the different orientation of the Earth's magnetosphere respect with XMM-Newton's orbit, confirming the solar origin of the soft proton population. The Equator-S proton spectrum at 70000 km altitude is used for the proton population entering the optics, where a combined multiple and Firsov scattering is used as physics interaction.
If the thick filter is used, the soft protons in the 30-70 keV energy range are the main contributors to the simulated spectrum below 10 keV. We are able to reproduce the proton vignetting observed in real data-sets, with a $\sim50\%$ decrease from the inner to the outer region, but a maximum flux of $\sim0.01$ counts cm$^{-2}$ s$^{-1}$ keV$^{-1}$ is obtained below 10 keV, about 5 times lower than the EPIC/MOS detection and 100 times lower than the EPIC/pn one. Given the high variability of the flare intensity, we conclude that an average spectrum, based on the analysis of a full season of soft proton events is required to compare Monte Carlo simulations with real events.
\end{abstract}

\keywords{Geant4, XMM-Newton, X-ray telescopes, background minimization, soft protons}

\section{INTRODUCTION}
\label{sec:intro}  

The Earth magnetic field confines the interplanetary energetic plasma (mainly protons and electrons) in two distinct radiation belts (the Van Allen belts). The inner one, more compact and mostly filled with protons, extends from about 700 to 10$^{4}$ km. The outer one, recently discovered to occasionally split into a third radiation ring\cite{2013Sci...340..186B}, is mainly constituted by high energy electrons and stretches up to $6\times10^{4}$ km. 
The trapped particles can pose a significant radiation threat to the on-board electronic systems, depending on the telescope orbit and technology.
Launched in July and December 1999 respectively, the NASA Chandra and the ESA XMM-Newton X-ray missions represented a cornerstone in the X-ray astronomical exploration, thanks to their unprecedented sensitivity and angular resolution below 10-15 keV. They are currently operating in a high eccentricity elliptical orbit that, reaching extreme apogees  ($>10^{5}$ km), allows them to spend most of the operating time beyond the radiation belts and without the frequent interruption of the Earth's shadowing. On the contrary, in the perigee passage the spacecraft altitude is in the range $(5-10)\times10^{3}$ km, forcing the telescopes to cross the radiation belts each orbit. 
\\
Both telescopes carry Wolter type-I mirrors to focus X-rays through grazing angle reflection to the detection plane, which is shielded by heavy absorbers outside the field of view to block penetrating radiation (mainly cosmic rays) to reach the focal plane.  Electrons in the radiation belts and in the outer regions can also be scattered by the X-ray optics and funnelled to the focal plane. For this reason, X-ray telescopes on board Chandra, XMM-Newton and Swift are equipped with magnetic diverters that deflect their path outside the detection plane (see e.g Ref. \citenum{wil00}). What came unexpected after one month of operation of the Chandra mission was that also low energy protons entering the field of view at grazing angles can be reflected too by the X-ray mirror shells and reach the focal plane. With energies below 100-300 keV, these so-called "soft protons" can cause serious damages to the electronic system and the overall mission performance. 
\\
XMM-Newton's filter wheel completely blocks the European Photon Imaging Camera (EPIC) field of view when crossing the radiation belts protecting the detectors from damage. However, the soft protons populating the outer magnetosphere, although with a $\sim10^3$ lower flux\cite{2000SPIE.4140...99O}, including the magnetotail, the magnetosheath and the solar wind, both in the form of a steady flux and violent Coronal Mass Ejections (CMEs), can produce photon-like deposits in the read-out system that increase the X-ray residual background level and even threaten the observation itself \cite{bri00}.
The performance of future X-ray focusing telescopes operating outside the radiation belts (e.g., the ESA Athena\cite{2013arXiv1306.2307N} mission and the eROSITA\cite{2014SPIE.9144E..1TP} instrument on-board the Russian Spektr-RG observatory, both to be placed in L2 orbit) might also be affected by soft proton contaminations. 
Monte Carlo simulations are mandatory to evaluate the impact of such events and accordingly design shielding solutions (e.g., a magnetic diverter) without limiting the sensitivity of the instruments, and require the accurate characterization of (i) the distribution of the soft proton population, (ii) the mirror-proton physics interaction at play, and (iii) the effect on the focal plane. Despite many solutions proposed so far to explain the physics interaction behind the soft proton grazing angle scattering, the difficulty in obtaining accurate laboratory measurements\cite{2015ExA....39..343D} prevents ray-tracing codes from implementing physically-reliable models. 
After presenting an overview of the XMM-Newton soft proton flare behaviour (Sec. \ref{sec:obs}), we test the current knowledge of the input proton population and the physics interaction (Sec. \ref{sec:physics}) by simulating the soft proton interaction with a simplified model of the X-ray mirror module and the EPIC/pn camera. The final aim is comparing for the first time the simulated X-ray spectrum with a real on-flight XMM-Newton observation of a soft proton flare (Sec. \ref{sec:sim}).
In the present work, we use Geant4 9.1 - current release at the time of writing being 10.2 - to cross-check our results with the ESA Geant4 simulation campaign (see e.g. [\citenum{nar02a}] outcome after the launch of Chandra.

\section{XMM-NEWTON OBSERVATIONS}\label{sec:obs}
  \begin{figure} [h!]
   \begin{center}
   \begin{tabular}{c} 
   \includegraphics[trim={0 0.8cm 0 0},clip,width=0.45\textwidth]{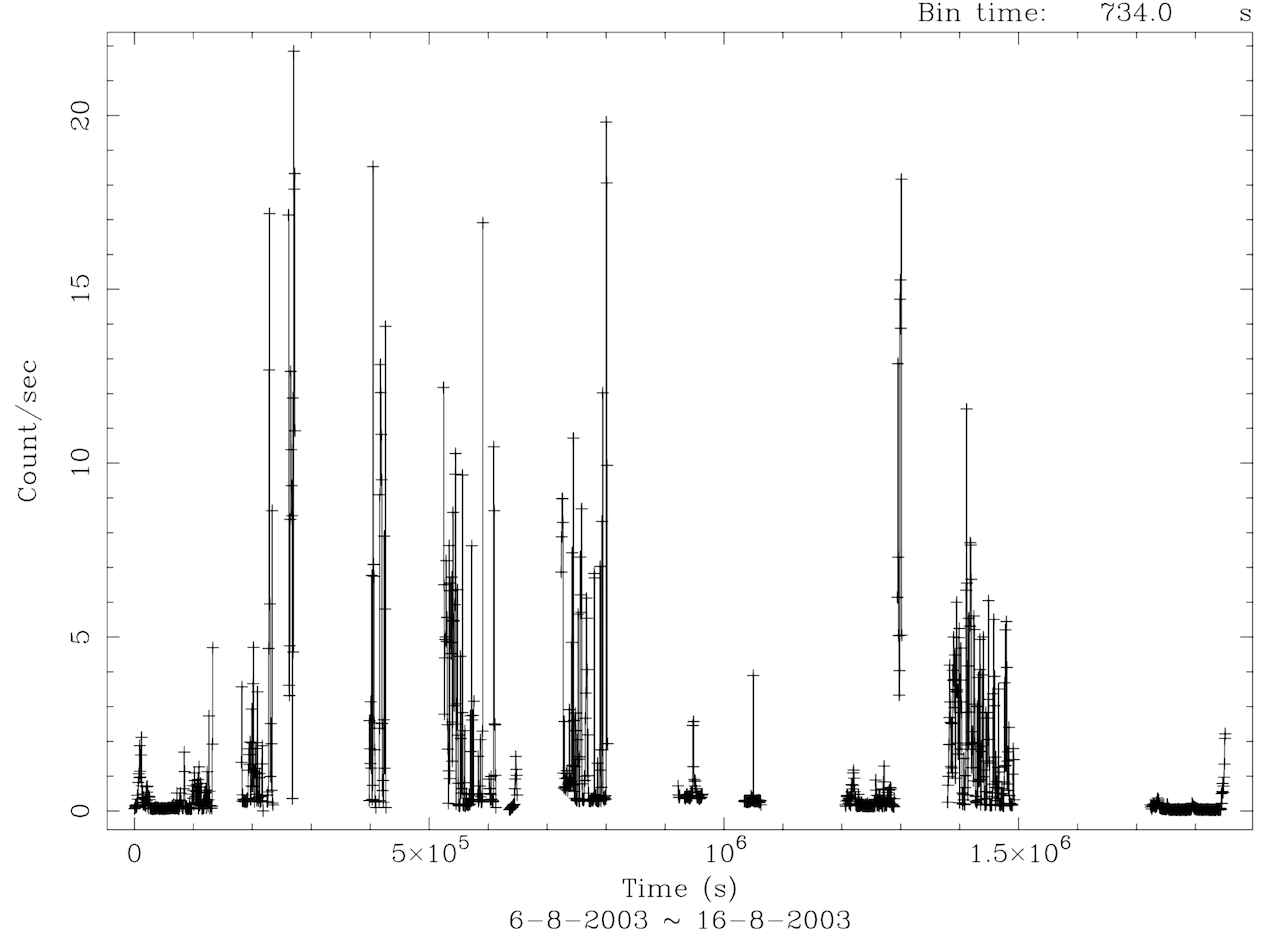}
   \includegraphics[trim={0 0.9cm 0 0},clip,width=0.45\textwidth]{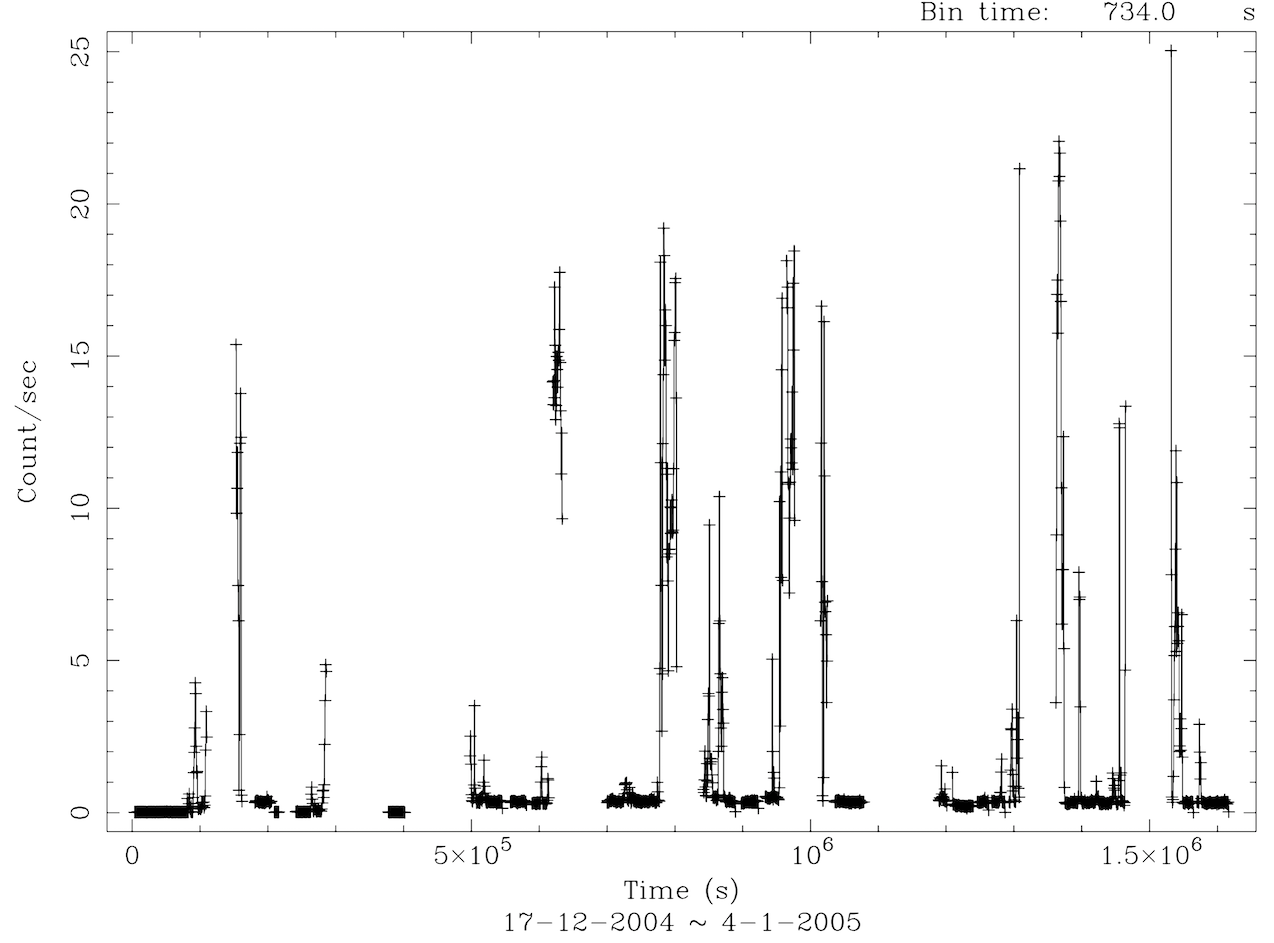}\\
   \includegraphics[trim={0 1.5cm 0 0},clip,width=0.45\textwidth]{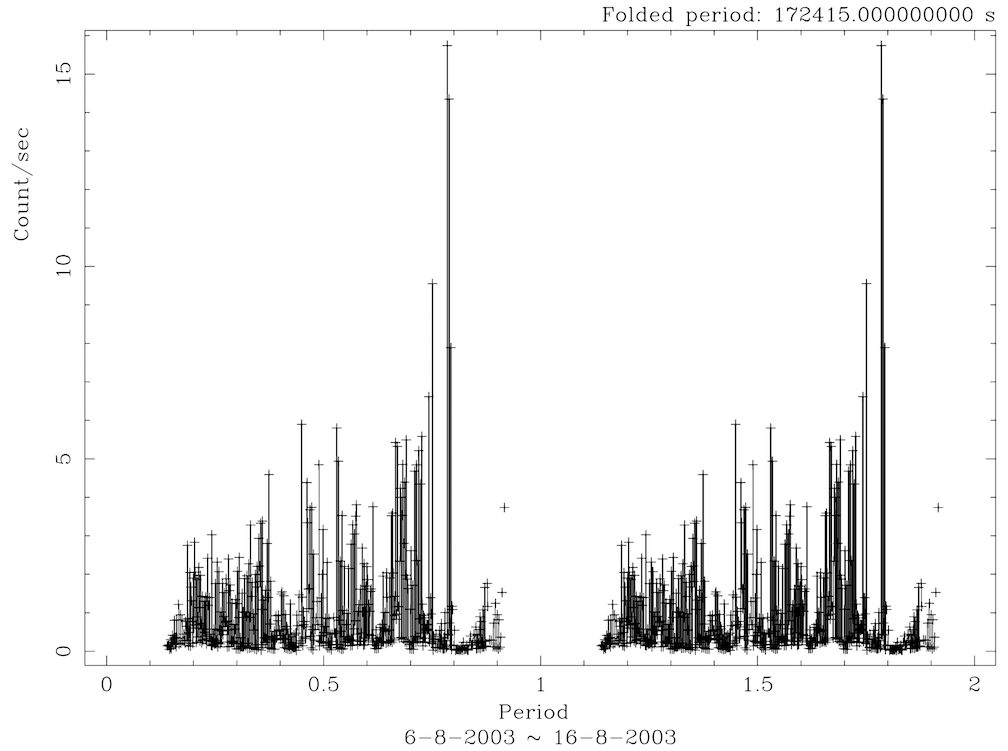}
   \includegraphics[trim={0 1.5cm 0 0},clip,width=0.45\textwidth]{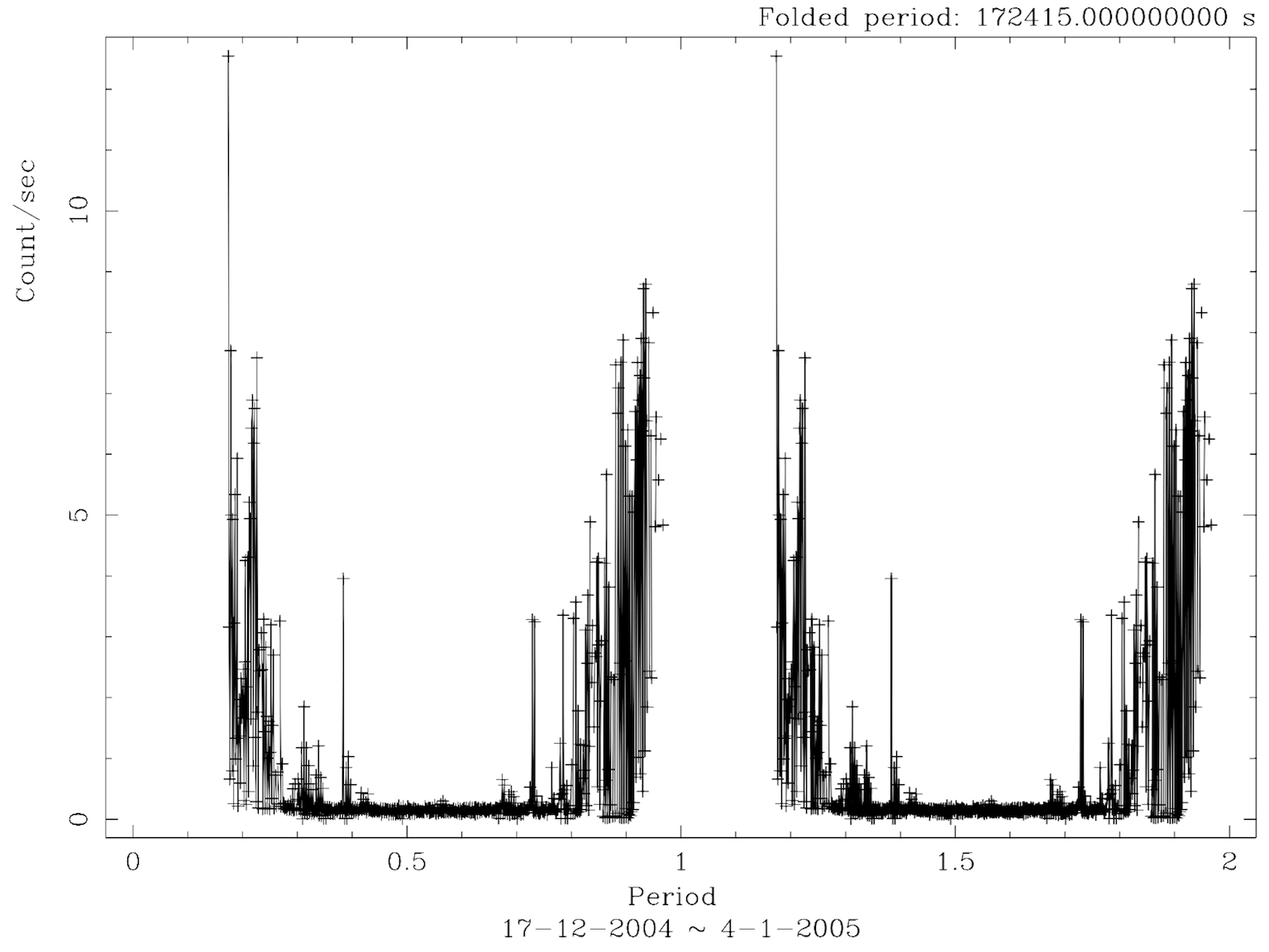}
   \end{tabular}
   \end{center}
   \caption 
   { \label{fig:softp_xmm}Nominal (top panels) and folded (bottom panels) light curves, using the $\sim48$ hr orbital period, of the $>10$ keV counts detected by the EPIC/pn instrument along 10 consecutive orbits in August 2003 (left) and December 2004 - January 2005 (right). A \emph{FLAG}=0 and \emph{PATTERN}=0 data selection is applied to the event list. The count rates in the y-axis are normalized accounting for the value of X-ray effective area.}
   \end{figure} 
After the first eight unprotected perigee passages on September 1999\cite{ode00}, the Chandra Advanced CCD Imaging Spectrometer (ACIS\cite{2003SPIE.4851...28G}) Front-Illuminated (FI) CCDs suffered a degradation of a $\sim100$ factor of the Charge Transfer Efficiency (CTE) with consequent loss of energy and image resolution\cite{2000SPIE.4140...99O}. On the contrary, no degradation was found for the Back-Illuminated (BI) CCDs. This fact, coupled with the lack of degradation when the ACIS was moved out of the field of view during the radiation belt passage, and the type of CTE damage, led to identify low energy (100 - 300 keV) proton scattering with the mirror shells \cite{kol00} as the main source of non ionizing energy losses and the production of permanent lattice displacement damage in the sensitive area\cite{pri00}.
The ACIS damage was soon minimized by moving the CCDs out of the proton sight line and switching off the instrumentation during the perigee passage. In addition, automatic instrument saving procedures are triggered by the on-board radiation monitor and real-time space environment data\cite{2007SPIE.6686E..03O}.
\\
The XMM-Newton EPIC instrument\cite{1996SPIE.2808..402V} is the telescope X-ray imager and spectrometer covering the 0.1 - 15 keV energy range, and it consists of three cameras: two MOS and one pn CCD. In addition to the focal plane instrumentation, two of the three telescopes are equipped with Reflection Grating Spectrometers (RGS\cite{2004SPIE.5501...32D}).
About $50\%$ of the damaging proton flux of the ACIS, within the same energy range, was expected for the XMM-Newton EPIC/MOS \cite{nar02a}, which is a FI instrument but is partially shaded by the RGS gratings, while no degradation was foreseen for the BI EPIC/pn\cite{nar02a}. Extensive simulation campaigns were able to prove that the XMM-Newton instruments could be sufficiently protected by closing the filter wheel when crossing the radiation belt. The minimum required satellite elevation to conduct science observations is currently 46000 km, and the instruments are switched--off in case of intense solar activity, continuously analysed by the radiation monitor that is sensitive to proton energies $>1.3$ MeV\footnote{http://xmm2.esac.esa.int/external/xmm\_obs\_info/radmon/radmon\_details/index.php}.
\\
Unfortunately above the radiation belt limit, where the instruments are fully operative, the soft proton funnelling is still observed by the XMM-Newton detectors in the form of a sudden increase in the background level, with events that can hardly be disentangled from the X-ray events and consequently cannot be rejected on board. Such soft proton flaring events can prevail over the quiescent background level up to 1000\%, affecting the 30-40\% of XMM-Newton observing time \cite{2007A&A...464.1155C}. Soft proton flares are extremely unpredictable in duration, lasting from $\sim100$ s to hours \cite{del04}, and generate an average count rate, in all the three CCDs, of 2-2.5 prot. cm$^{-2}$ s$^{-1}$ \cite{lum02}. Several studies (see e.g., Ref. \citenum{2007A&A...464.1155C}) have proven the solar origin of this damaging background. Missions operating in low Earth orbit (e.g., Swift, Suzaku) do not suffer from soft proton flares, thanks to the geomagnetic shield.
\\
The X-ray spectrum generated by proton scattering, spanning all over the EPIC operative energy range\cite{2005A&A...443..721P} from $\sim0.5$ keV, is characterized by a continuous, featureless emission, well fitted by a broken power-law, with a variable break energy ($\sim3$ keV for MOS according to Ref. \citenum{kun08} and $\sim20$ keV for pn according to Ref. \citenum{ken00}. In contrats, only 8\% of the Chandra observation time is affected by this background components, owing mainly to the lower collecting area of the mirror module and the smaller detector size.
\\
Higher is the X-ray energy, the lower is the critical reflection angle on the mirror shells. This effect, combined with the decrease in quantum efficiency ($\sim50\%$ from 10 to 15 keV\cite{2001A&A...365L..18S}) makes the observation of an X-ray source photon very unlikely above 10 keV.
A key evidence of a soft proton induced event is the XMM-Newton detection, in many cases, of emission well above 10 keV\cite{lum02}, where the effective area to X-ray scattering starts to decrease and no photons from the scientific target should be expectedly detected. Fot this reason, we explored the phenomenological behavior of soft proton flares along XMM-Newton's orbit by studying the distribution of $>10$ keV event in the EPIC/pn instrument. 
  \begin{figure} [t]
   \begin{center}
   \begin{tabular}{c} 
   \includegraphics[width=0.8\textwidth]{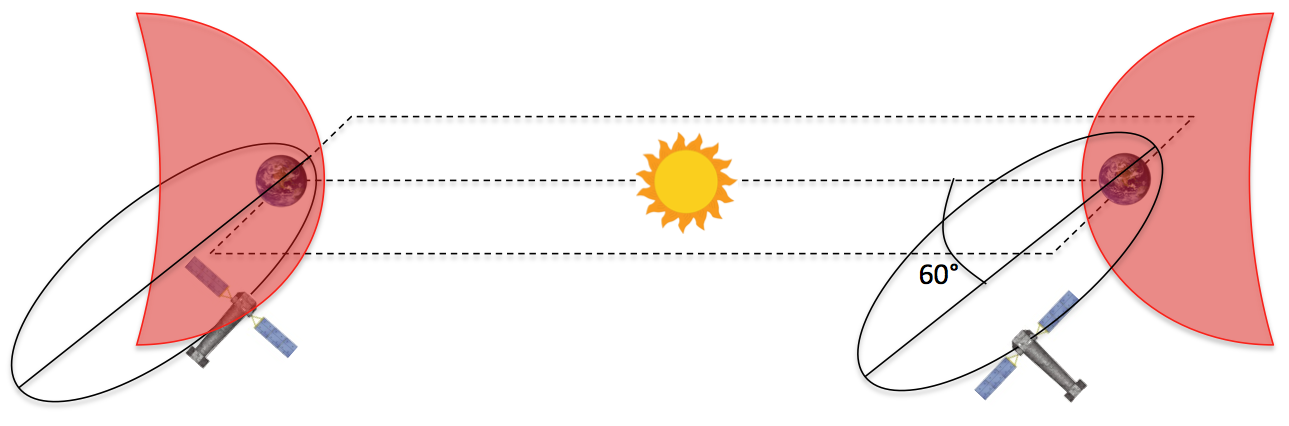}
   \end{tabular}
   \end{center}
   \caption 
   { \label{fig:orbit}A schematic of XMM-Newton orbit orientation respect with the magnetosphere for two opposite Earth's positions towards the Sun: the northern emisphere summer, shown on the left, and winter, shown on the right. The orbit inclination angle of $60$ deg. (May 2010 update) is also overlaid.}
   \end{figure}    
\subsection{Data reduction and event selection}
The EPIC/pn data set has been reduced and analyzed following the standard procedures as adviced by the XMM-Newton SOC using SAS\footnote{http://www.cosmos.esa.int/web/xmm-newton/what-is-sas}. The observation lasts 10 consecutive telescope orbits ($\sim2$ Ms) for two different seasons: 1-16/08/2003 (summer) and 17/12/2004 - 4/01/2005 (winter). In the two seasons, the Earth's position with respect to the Sun, and therefore with respect to the geomagnetic tail, is opposite.
We used the \emph{evselect} tool to extract from the event list all the counts above 10 keV and create a time serie of the count rate, in counts s$^{-1}$ as a function of time, using a time binning of 734 s.
In addition to the energy selection, allowing us to avoid the contamination from X-ray sources, we used the \emph{FLAG} and \emph{PATTERN} data markers to better select soft proton events. The \emph{FLAG} keyword, indicating the data quality of the event, has been set to 0 to exclude all the events at the CCD border or bad pixels, that would be excluded in a scientific observation. The \emph{PATTERN} keyword, describing the distribution of the electronic cloud on the detector by counting the number of pixels triggering the same event, has also been set to 0. As reported by [\citenum{del04}], most of the non X-ray background of XMM-Newton only involves single pixel events, as expected because of the collisional nature of its interactions.
\\
The light curves have been extracted using the \emph{lcurve} data analysis tool and epoch folded, i.e. dividing the time series by a period and summing all the intervals, using the \emph{efold} tool of the HEASARC \emph{Xanadu} package\footnote{https://heasarc.gsfc.nasa.gov/xanadu/xanadu.html}. The folding period is $\sim173$ ks (the XMM-Newton orbit) to highlight orbital effects on the flare detection. Figure \ref{fig:softp_xmm} shows the resulting nominal light curves (top panels) and folded (bottom panels) light curves for the summer (left panels) and winter (right panels) EPIC/pn observations.

\subsection{Periodicities and seasonal effect}

The blank areas in the nominal light curves of Fig. \ref{fig:softp_xmm} represent the passage through the radiation belts. Extreme intensity-variable peaks are found in both the two seasonal light curves. The epoch folded light curves start at the perigee passage. Flares are detected soon after the switching-on of the instruments and the filter wheel opening above the 46000 km altitude, but major differences are found if the summer plots are compared to the winter ones.
An observation in different seasons means that the Earth's position towards the Sun, hence the distribution of the magnetosphere and its tail, is different. A seasonal effect on the detected soft proton flares should be expected if low energy protons, after being emitted by the Sun and decelerated by the interaction with the Earth's geomagnetic field, stay trapped in the Earth's local space environment. The winter folded light curve (Figure \ref{fig:softp_xmm}, right bottom panel) shows an higher flare rate in proximity of the radiation belts, due to a residual trapped proton population, and then a sudden flare decrease at the apogee passage. A wealth of peaks is instead observed along the entire summer orbit. Although different solar events are surely affecting the two data sets, we think that local transient events cannot explain the different flare density detected in the summer/winter observations.
From the analysis of the Earth and satellite orbit on the soft proton flare detection we find that:
\begin{itemize}
\item soft proton events are also detected in the apogee region, well far from the inner radiation belt, with extreme intensities, regardless of the different observing period;
\item the higher density of count rate peaks along the entire orbit during summer could be explained by the different angle between the highly eccentric satellite orbit and the magnetosphere axis, with the telescope crossing a larger portion of trapped particle regions (see Figure \ref{fig:orbit}).
\end{itemize}
   \begin{figure} [h]
   \begin{center}
   \begin{tabular}{c} 
   \includegraphics[height=8cm]{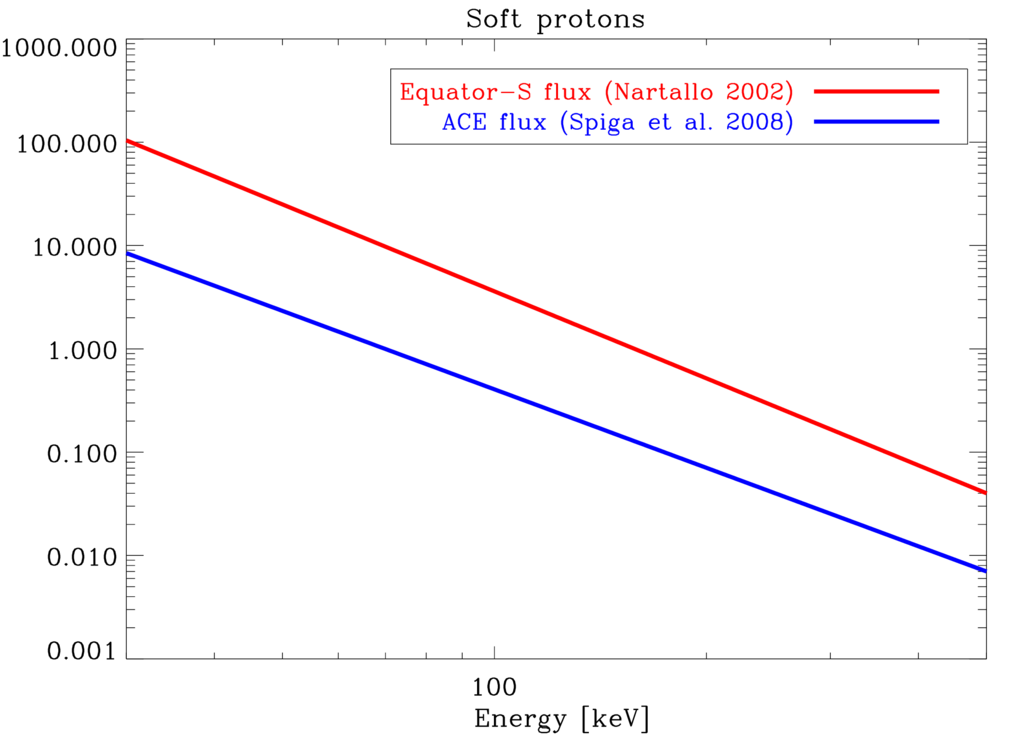}
	\end{tabular}
	\end{center}
   \caption[example] 
   { \label{fig:softp_spe} 
Differential flux, in prot. cm$^{-2}$ s$^{-1}$ keV$^{-1}$ sr$^{-1}$, of the low energy protons as detected by the Equator-S (red line) at 70000 km altitude and by the ACE mission (blue line) in L1.}
   \end{figure} 
\subsection{A model for the soft proton population}
Since the discovery of the soft proton population is relatively recent, models for their distribution are still poor, and are complicated by the fact that their energy peaks below the sensitive range of the radiation monitor on board XMM-Newton\cite{ken00}. At the same time, their intensity is highly variable, with a seasonal effect as the geotail (the magnetosphere night side) sweeps around the Earth, as proven in the previous section, and a short term variability due to the solar and geomagnetic activity \cite{bla00}. The accurate modeling of the energy distribution of the protons reflected by X-ray mirrors is the first major issue when simulating a soft proton flare observation.
In Ref. [\citenum{nar02b}] a proton spectrum from 30 keV to 1 MeV is extracted from the observation of the Equator-S\cite{2000AdSpR..25.1277H} satellite, a German small mission launched in 1997, and aborted after 5 months, to study the Earth's equatorial magnetosphere out to a distance of 67000 km, where XMM-Newton operates. They fit the differential flux detected by the Energetic Particle Instrument (EPI) with a power law $\propto \rm E^{-2.8}$, with a measured proton flux of $\sim3$ prot. cm$^{-2}$ s$^{-1}$ keV$^{-1}$ sr$^{-1}$ at 100 keV. It is plotted in red in Figure \ref{fig:softp_spe}. The Electron, Proton, and Alpha Monitor (EPAM\cite{gol98}) on board the NASA Advanced Composition Explorer (ACE) satellite, launched in 1997 and still operating, studies ions ($>50$ keV) and electrons ($>40$ keV) of solar and galactic origin, acting as a near real-time solar wind monitor. It orbits in L1 ($\sim1.5$ million km from the Earth at the Sun side), and it is able to provide an accurate measurement of the proton population in the farest regions. The ACE/EPAM differential proton spectrum is shown in Fig. \ref{fig:softp_spe} (blue line) as reported by Ref. [\citenum{spi08}], with a power-law energy distribution and a photon index around 2.
The average proton flux at 70000 km altitude is about an order of magnitude higher than the pure solar wind proton contribution: following a conservative approach, we use the Equator-S model as input for the simulation described in Sec. \ref{sec:sim}.
\section{PHYSICS INTERACTION}\label{sec:physics}
Extensive Geant4\cite{g4_1, g4_2} Monte Carlo (MC) simulations have been performed since the initial damage suffered by Chandra to characterize the nature of the soft proton scattering process with X-ray mirror shells.
The Geant4 Monte Carlo toolkit is an open source, C++ based, particle transport code, initally developed by CERN for the simulation of high energy experiments at particle accelerators and then extended to “lower” energy ranges, i.e. the X and Gamma-ray domain. Geant4 has become the standard tool used by many space agencies (e.g., ESA, NASA) in the simulation of the background and instrument performance of all major X-ray space telescopes (e.g., Chandra, XMM-Newton, Suzaku, Athena, eROSITA).
\\
The use of accurate models that are able to reproduce the correct proton energy, spatial, and angular distribution at the optics exit is mandatory for the simulation and consequent design of any future X-ray mission facing soft proton background events. At the time of writing, the following physics process have been tested in MC simulations:
\begin{itemize}
\item Geant4 simulations of soft proton interaction with Chandra and XMM-Newton optics carried by [\citenum{nar01}] use multiple Coulomb - or Rutherford - scattering as the dominant physics process. In the Geant4 9.1 release\footnote{http://geant4.web.cern.ch/geant4/support/ReleaseNotes4.9.1.html}, the multiple scattering class uses the Lewis theory, including both the moment of the spatial and the angular distribution after a step, instead of the Moliere formalism, as the Geant4 physics list reported in [\citenum{nar01}]. Each collision in this formalism is independent of the others, and the resulting angle and energy distribution at the exit of mirror surface is the sum of small, random changes of the proton momentum and path. 
\item as reported by [\citenum{dw03}], the assumption of independent collisions in the multiple scattering model is only valid if the proton travels within the bulk material, which is only possible for large incidence angles. For protons reaching the surface at grazing angles, as the case with X-ray mirrors, the interaction with the electron plasma cloud at the surface causes the near-specular scattering\cite{2001PhRvA..63e2902S} with no collision with atomic nuclei and weak dependence on the energy and angle of incidence. In the limit of zero scattering energy loss and for small incident angles, the authors describe the scattering response using the Firsov formula\cite{firsov}. Since the function peaks at specular reflections, this model produces lower scattering angles than multiple scattering This fact, coupled with smaller energy losses that shift the damaging proton population at lower energies, produces higher fluxes of protons at the detector surface. The Geant4 implementation of the Firsov model by [\citenum{lei04}] has been adapted here to the Geant4 9.1 release. The same code has also been used for the evaluation of the protons and electrons scattering through the IXO X-ray optics\cite{san09}. 
\end{itemize}
If the proton is treated as a de Broglie wave in conjunction with an index of refection less than one, the work of [\citenum{asc07}] predicts a total reflection of protons with energies below few MeV and an incident angle less than 1 deg, producing a scattering distribution of 1 deg. or wider depending on the mirror microroughness. No Geant4 simulations have been performed up to date to test the wave-like model.
  \begin{figure}[h!]
   \centering
   \includegraphics[width=0.7\textwidth]{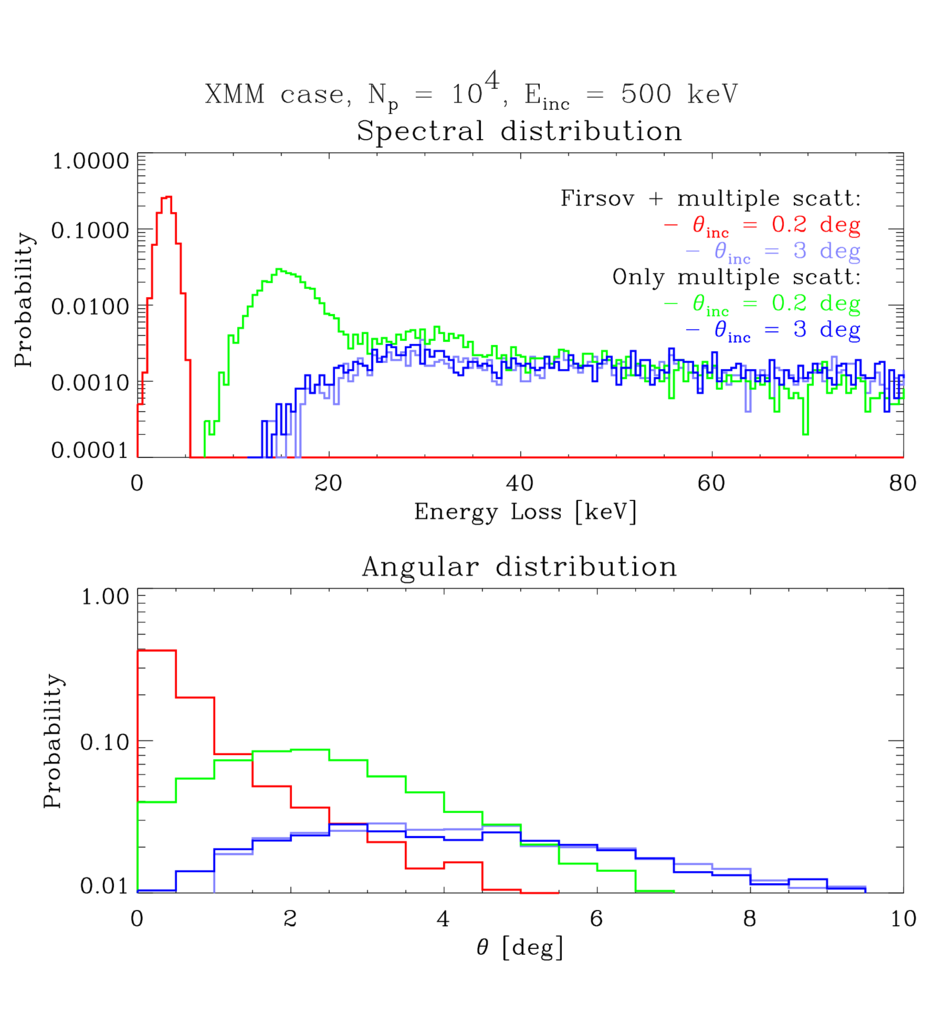}
   \caption{\label{fig:fir}Probability distribution of the energy loss (top panel) and angular distribution (bottom panel), from the telescope axis, of 500 keV protons after the interaction with an XMM mirror shell. The red and light blue lines refer to the activation of the multiple and Firsov scattering for an incident angle of 0.2 and 3 deg., while the green and blue lines only use multiple scattering, for the same incident angles as the Firsov case.}
    \end{figure}
\\
Proton reflectivity measurements were performed by [\citenum{ras99}] on XMM-Newton grating and mirror samples soon after the discovery of the ACIS damage using 0.3 and 1.3 MeV incident protons. Despite the claim in [\citenum{nar01}] that Geant4 simulations using multiple scattering reproduce well the experiment, more than a factor 10 discrepancies\cite{nar02a} are found for small ($\leq1$ deg.) incident angles, especially at 0.3 MeV incident proton energy. In addition, no information is available for the energy distribution of the scattered proton obtained in the experiment.
Recent measurements of proton scattering at 250 keV, 500 keV, and 1 MeV using spare mirror shells of the eROSITA instrument\cite{2015ExA....39..343D} find a scattering efficiency around a factor 2 higher at 500 keV and a factor 3-4 higher at 1.3 MeV, but the comparison could be biased by the different method used in computing the incident flux. The resulting scattering distribution on the eROSITA shell is well reproduced by the Firsov model, but larger energy losses than a pure elastic scattering are found.
\\
In the present Geant4 simulation we decided to combine both the Firsov and the multiple scattering,
Following the work of [\citenum{lei04}], the Firsov model is applied for incident angles below 1 degree, and the standard Coulomb collisions start above this limit. 
Since laboratory measurements of grazing proton scattering on Aluminum surfaces\cite{1997NIMPB.125..124W} show a most probable energy loss around 3 keV in the 30-710 keV proton energy range, weakly dependent on the proton incident energy and angle, and their result is consistent with energy losses obtained using different mirror materials\cite{1987PhRvB..36....7K}, a constant energy loss of 3 keV is applied to each simulated Firsov interaction.  
\\
A beam of 500 keV proton is emitted towards a slab composed by a 1 mm thick Nickel (Ni) coated by a 50 nm Gold (Au) layer, similar to XMM-Newton X-ray shells, for two incident angles with respect to the slab surface: $0.2$ and $3$ deg.
In Figure \ref{fig:fir} the energy loss (top panel) and the scattering angle (bottom panel) probability distribution are plotted with and without the addition of the Firsov scattering, the latter being activated only for incident angles lower than 1 degrees.
For an incident angle of 3 deg., both cases give the same result, because the Firsov scattering is not effective: the proton energy loss is equally distributed above 20 keV, and a broad scattering angular distribution is obtained up to 10 degrees. For an incident angle of 0.2 deg., the multiple scattering induces losses at $\sim15$ keV but still with a broad tail, while the exit proton distribution is more collimated. However, this result is far from the findings out of the Firsov scattering, inducing a lower energy loss ($\sim3$ keV) and a proton deflection at angles that are closer to the specular direction. The ability to reproduce the scattering probability of [\citenum{lei04}] validates our simulation set-up.
 
\section{BOGEMMS SIMULATION}\label{sec:sim}
The Bologna Geant4 Multi-Mission Simulator (BoGEMMS\cite{2012SPIE.8453E..35B}) is a modular and parameterized tool for the evaluation of the instrument performance (e.g., background induced radiation, detection efficiency, angular resolution) of X-ray and Gamma-ray space telescopes. The software project includes both the Geant4-based simulator architecture and the filtering/analysis suite. While exploting all the features of the Geant4 toolkit (e.g., the 3D geometry modeling, the tracking code, the physics lists), BoGEMMS adds an astronomer-friendly interface to easily modify at run-time the simulation set-up and obtaining as output an event list, in NASA FITS format\cite{FITS}, to produce high level scientific analysis products (e.g., X-ray spectra, count maps).
It has been used for the study of the shielding optimization of the Simbol-X and the New Hard X-ray Mission (NHXM) telescopes, and the overall study of the X-ray background radiation in Low Earth Orbit\cite{2012SPIE.8453E..31F}. We use the BoGEMMS framework to simulate the soft proton funnelling into XMM-Newton optics and their impact on the EPIC/pn X-ray spectrum.
 \begin{figure}[h!]
   \centering
   \includegraphics[trim={0 1.1cm 0 0},clip, width=0.5\textwidth]{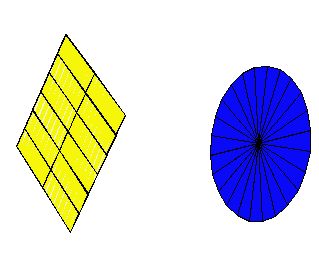}
   \hspace{2cm}
    \includegraphics[width=0.3\textwidth]{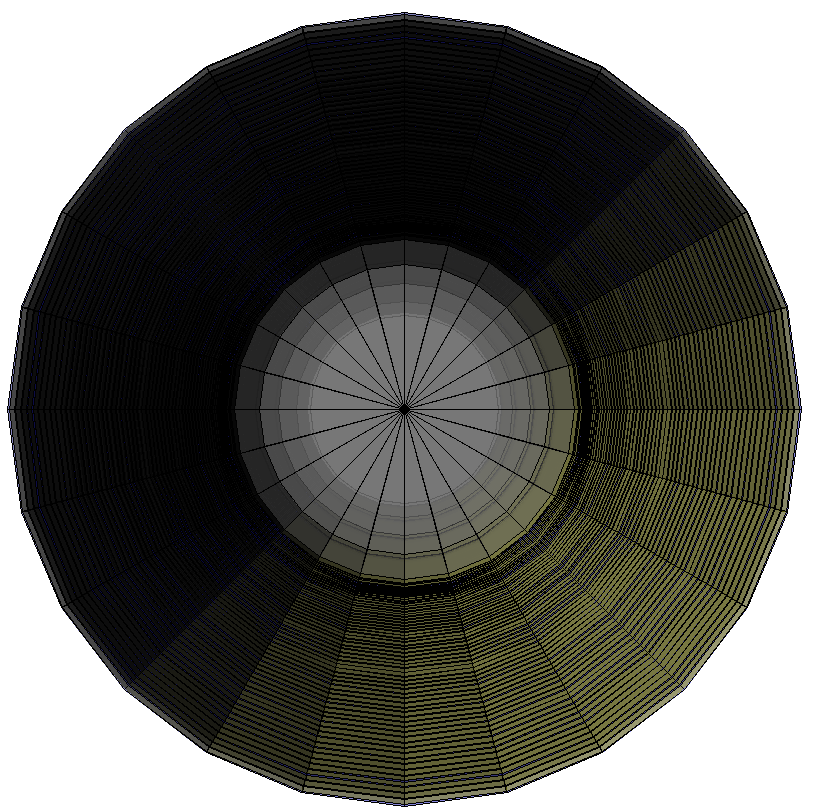}\\
    \vspace{1cm}

   \caption{\label{fig:xmm_g4}The XMM-Newton simplified Geant4 model - not in scale - composed by the EPIC/pn detector (in yellow on the left), the thick filter in front of it (in blue at the center), and the X-ray mirror module (on the right). }
    \end{figure}
A simplified model of the XMM-Newton mirror module has been developed (Figure \ref{fig:xmm_g4}), using as example of Wolter-I X-ray shell the \textit{XrayTel} exercise provided by the Geant4 user support team.
The simulated mirror module is composed by 58 shells, made of 1 mm thick Ni layers covered by a 50 nm Au coating. Four conical sections are assembled together to approximate the paraboloid-hyperboloid shape.
With a total height of 60 cm, the shell radius at the paraboloid-hyperboloid intersection ranges from $\sim150$ to $\sim350$ mm. The simplified EPIC/pn geometry only comprises the Silicon sensitive surface\cite{str00}: $6\times2$ modules, of dimension $30\times9.6$ mm, divided into $200\times64$ squared pixels, with a side of 150 $\mu$m and a thickness of 0.3 mm. The modules are centrally divided by a gap of 41 $\mu$m, and separated by a 198 $\mu$m dead area. A distance of 7.2 m is placed between the focal plane and the exit pupil of the mirror module. A thin, a medium, and a thick filter are mounted on the XMM-Newton filter wheel to protect the EPIC CCDs from the infrared, the visible, and the ultraviolect light of the science target and avoid contamination. We decided to reproduce the thick optical blocking filter, represented in the simulation without the supporting structure, because the reference soft proton flare observations are performed with the thick filter on (see Sec. \ref{sec:res}). Since the presence of the filter reduces the proton energy at the detector surface, it must be taken into account when comparing the results with a real observation. The filter composition and thickness are listed in Table \ref{tab:filt} and are taken from [\citenum{tie07}].
\begin{table}[h]
\begin{center}
\begin{tabular}{c|c|c}
\multicolumn{3}{c}{\textsc{The XMM-Newton Thick Filter}}\\
\hline
\hline
\multirow{2}{*}{Layer} & \multirow{2}{*}{Composition} & \multirow{2}{*}{Thickness [$\mu$m]}\\
 & & \\
\hline
1  & Polypropylene (PP)& 0.33\\
2  & Aluminum (Al)& 0.11\\
3  & Tin (Sn)& 0.045\\
\hline
\end{tabular}
\end{center}
\caption{\label{tab:filt}The composition and thickness of the XMM-Newton thick optical blocking filter. The layer number goes from the detector side towards the optical module.}
\end{table} 
\\
The Equator-S input proton differential flux (Sec. \ref{sec:obs}) is simulated from 30 to 500 keV as a power-law in the form:
\begin{equation}
\rm dF(E)/dE = 10^{6.15}\;E^{-2.80}\;\rm prot. \;cm^{-2}\;s^{-1}\;keV^{-1}\;sr^{-1}\;.
\end{equation}
Protons are emitted from an annular planar source (see Figure \ref{fig:old}, top-left panel), extending from $\sim15$ to $\sim35$ cm, placed at the entrance of the X-ray optics, covering exactly the entrance pupil of the mirror module to increase the simulation statistics. For each originating point on the annulus, protons are emitted isotropically within a cone coaxial to the telescope and variable angular aperture. From testing three different cone aperture angles, $\pm10$, $\pm5$, and $\pm2$ deg. the same final distribution on the focal plane is found (see Figure \ref{fig:no_real}, left panel), i.e. for off-set angles above 2 deg. the proton scattering efficiency drops off. 

\subsection{X-ray optics interaction effect on the proton population}
  \begin{figure}[h!]
   \centering
   \includegraphics[width=0.8\textwidth]{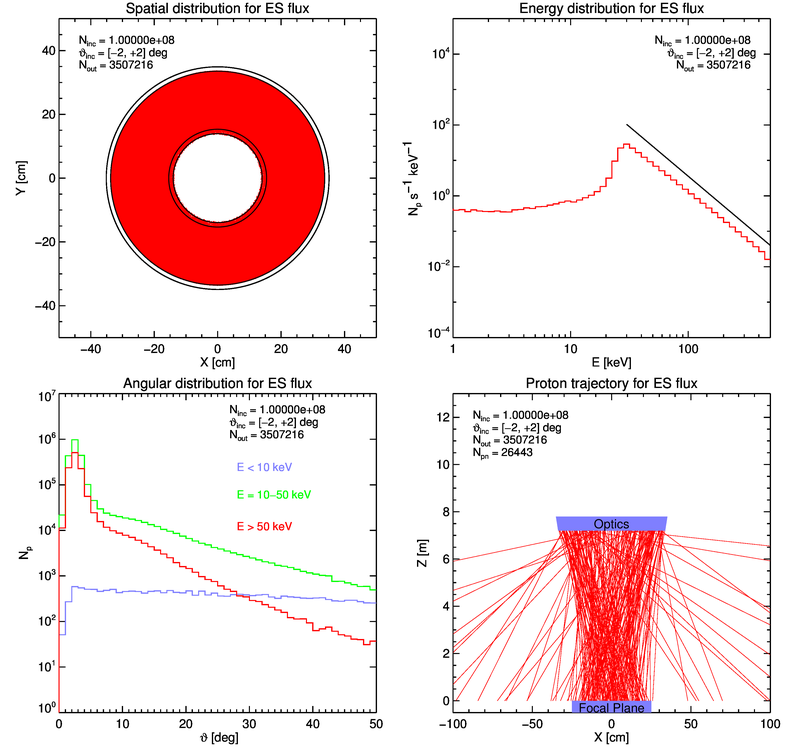}
   \caption{\label{fig:old} Spatial, angular and energy distribution of the soft protons after the interaction with the mirror module. \textit{Top-left}: spatial distribution of the protons, plotted as red points, exiting the X-ray shells; \textit{top-right}: spectral distribution of the proton at the optics exit (red line) compared to the initial Equator-S flux (black line); \textit{bottom-left}: angular distribution, with respect to the telescope axis, of the proton after the interaction for three energy ranges; \textit{bottom-right}: proton track visualization, in red in the X-Z plane, of the protons after their interaction with the optics.}
   \end{figure}
     \begin{figure}[h!]
   \centering
   \includegraphics[width=0.37\textwidth]{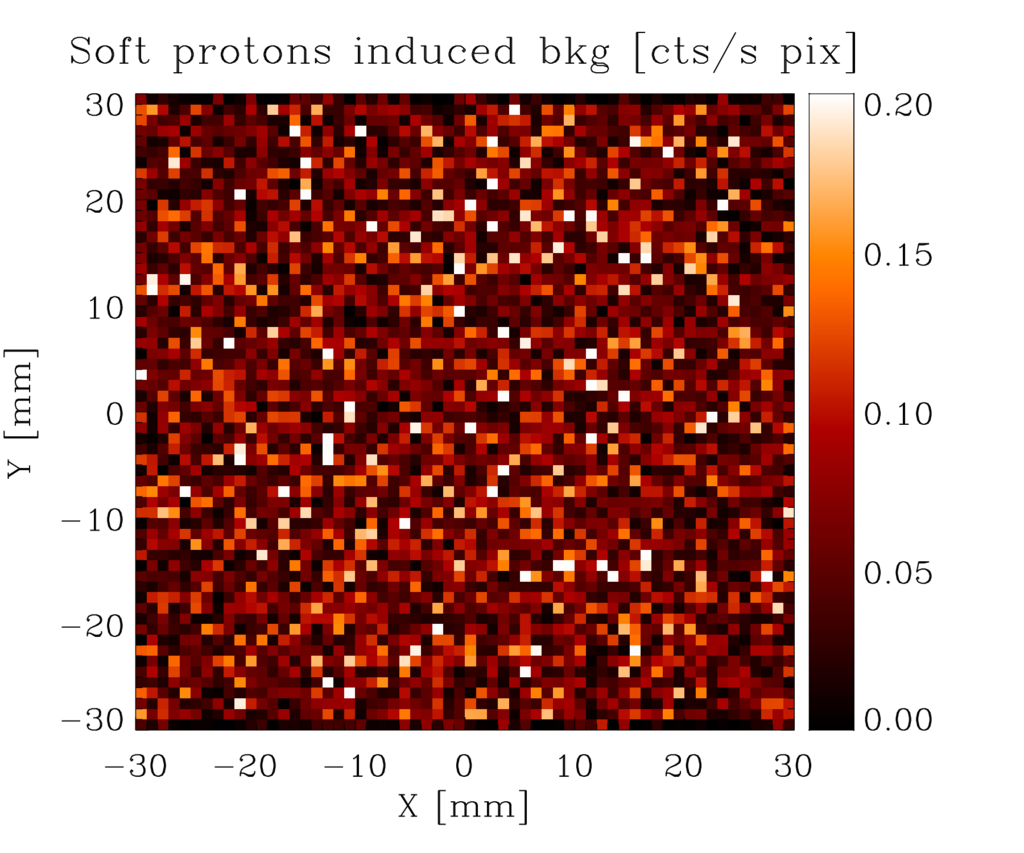}
   \includegraphics[trim={19cm 0 0 0},clip,width=0.29\textwidth]{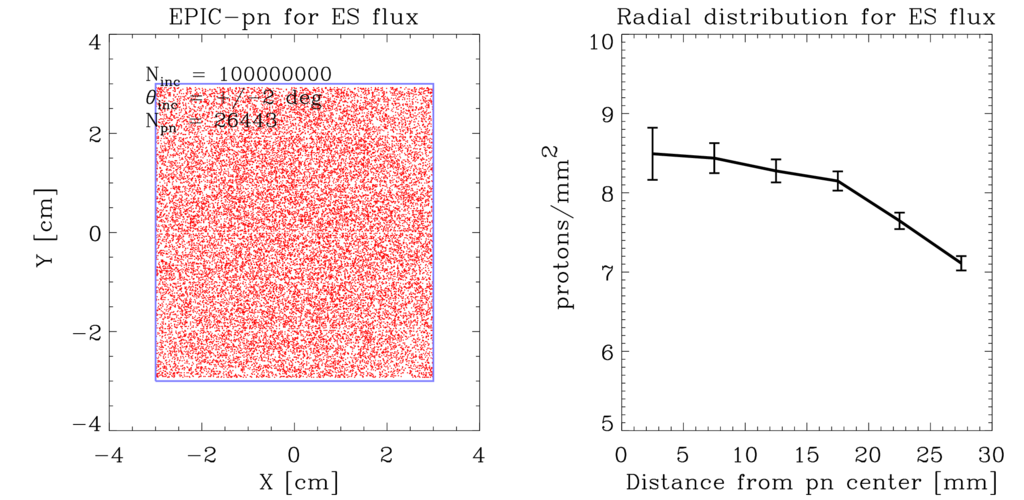}
   \includegraphics[trim={19cm 0 0 0},clip,width=0.29\textwidth]{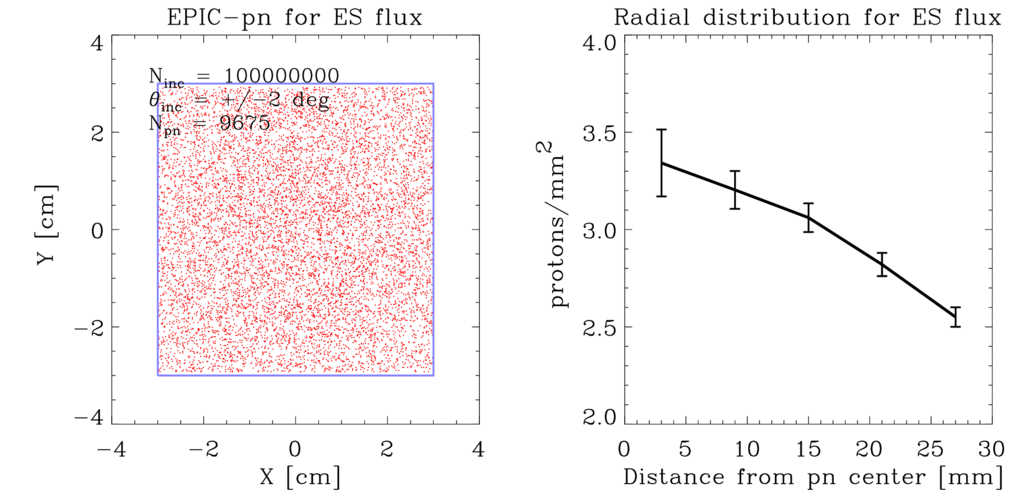}
   \caption{\label{fig:vign_1} The simulated binned counts map detected by the EPIC/pn (left panel) and the radial vignetting effect, for all the events (central panel) and E$<10$ keV events (right panel), of the soft proton intensity distribution.}
   \end{figure}
The energy, angular, and spatial distribution of the protons after the interaction with the X-ray optics, due to the combination of multiple and Firsov scattering depending on the incident angle, is presented in Figure \ref{fig:old}. The proton distribution at the exit of the X-ray mirror module is uniform (top-left panel) but only 3.5\% of the emitted population is able to cross the module (we do not count protons crossing the shells and exiting at the lateral side of the optics). 
The combination of Firsov interactions, inducing a constant 3 keV energy loss, and multiple scattering, with much larger energy losses, causes a shift toward softer energies of the primary distribution plus a long low energy tail covering the EPIC operative range (top-right panel). As expected, the protons populating the soft tail below 10 keV exhibit an uniform $\theta$ distribution from few degrees to 50 deg. because of multiple scatterings that cause large energy losses and reflections (bottom-left panel). On the contrary above 10 keV the distribution is peaked at $\sim2-3$ deg., then it decreases for larger angles, with a steeper decrease above 50 keV. Those are the protons that interacted by Firsov  scattering and present energies above 30 keV, the minimum simulated energy, at the optics exit. The bottom-right panel of Fig. \ref{fig:old} shows the tracks, in red, of the protons exiting the optics (top side) towards the focal plane (bottom side): the percentage of protons reaching the detector, a $6\times6$ cm area, is the 0.02\% of the emitted population and the 0.8\% of the protons funnelled by the X-ray shells, with many protons hitting a larger area of about $20\times20$ cm.
\\
Figure \ref{fig:vign_1} (left panel) shows the event density, in counts s$^{-1}$ pixel$^{-1}$, on the detector (a spatial binning of 1 mm instead of the real 0.15 mm EPIC/pn pixel side is used for visualization purposes). No specific patterns are visible from the counts map, but if we divide the EPIC/pn in annular regions from the center and count the surface event density, in protons mm$^{-2}$, a vignetted distribution is clearly detected (center panel of Fig. \ref{fig:vign_1}). The vignetting effect in X-ray focusing telescopes defines the intensity reduction, from the center to the edges of the detector, due to the lower effective area as the offset angle of the incoming X-ray photons increases. Soft proton flares also show a vignetted distribution on the focal plane of XMM-Newton (see e.g. [\citenum{kun08}]), an effect that contributed to the identification of the soft-proton funnelling interaction as the origin of such intense flares. The resulting vignetting from our simulations, if only $<10$ keV protons are selected (right panel of Fig. \ref{fig:vign_1}), is similar to what obtained by [\citenum{str00}] in a real observation, with a $\sim50\%$ decrease from the inner to the outer region.

\subsection{Resulting X-ray spectrum and comparison with real data}\label{sec:res}
\begin{figure}[h!]
   \centering
   \includegraphics[width=0.8\textwidth]{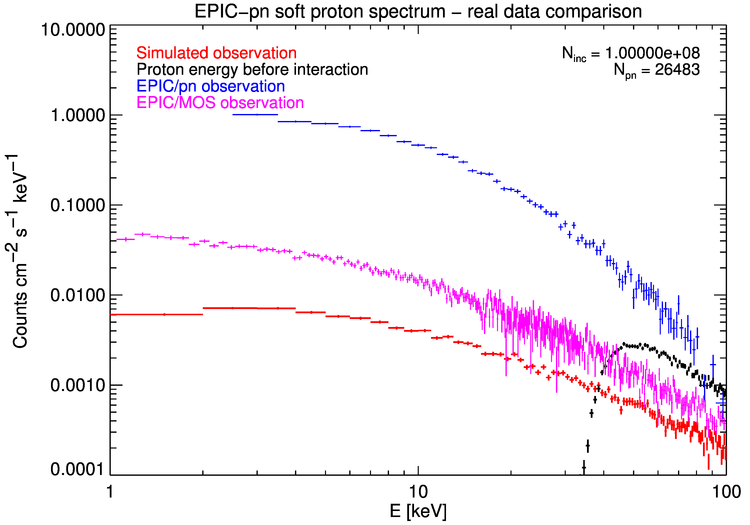}
\caption{\label{fig:prot_SPE}The simulated proton induced X-ray spectrum (in red) on the EPIC/pn, with the thick filter, obtained using as input the Equator-S proton flux and using a combination of Firsov and multiple scattering as physics interaction. The distribution of the initial energy of the protons depositing energy on the detector is plotted in black. The blue and purple spectra refer to the EPIC/pn and EPIC/MOS real soft proton flare observations.}
   \end{figure}

\begin{figure}[h!]
   \centering
   \includegraphics[width=0.49\textwidth]{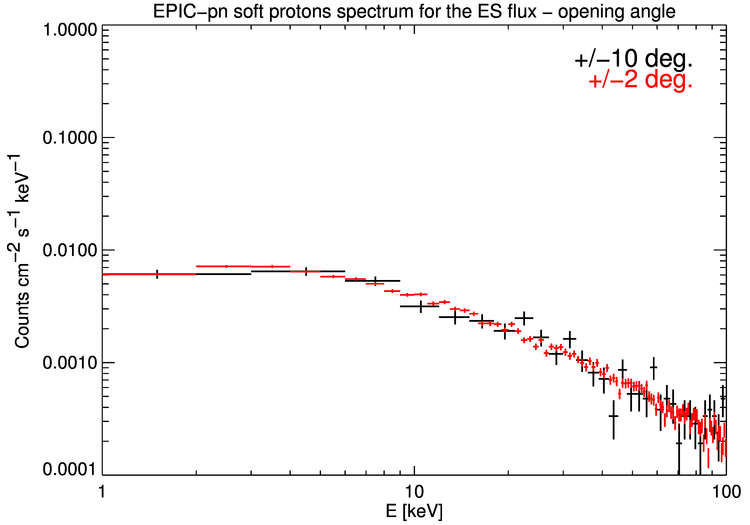}
   \includegraphics[width=0.49\textwidth]{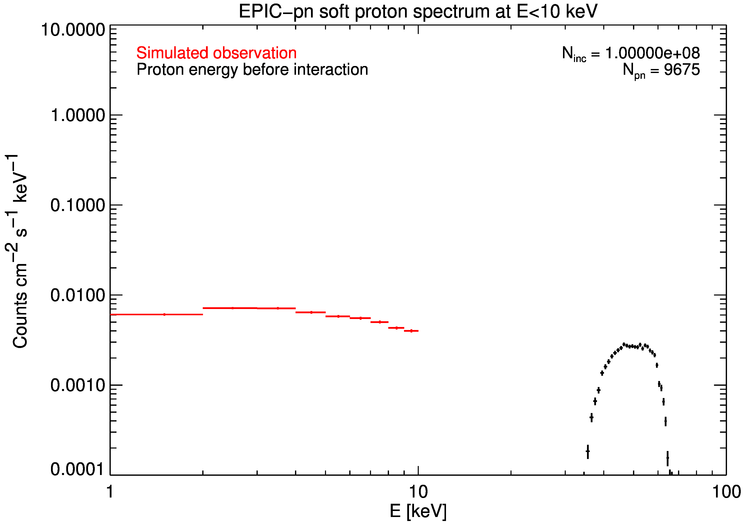}
   \caption{\label{fig:no_real} \textit{Left panel}: comparison of the simulated final X-ray spectrum for two aperture angle, $\pm10$ and $\pm2$ deg. of the emission cone at the mirror module entry; \textit{right panel}: the energy range of the protons at the optics entry if only E$<10$ keV counts on the EPIC/pn are selected.}
    \end{figure}

The number of simulated counts on the EPIC/pn are divided by the exposure, the detection area, and the bin energy content to compute the normalised rate, in counts cm$^{-2}$ s$^{-1}$ keV$^{-1}$, and to finally compare our result with real on-flight observations. The equivalent exposure time is computed by integrating the Equator-S model along the emitted energy range and solid angle. The resulting rate, in prot. s$^{-1}$, is used to divide the number of simulated protons ($10^8$ in this case) and obtain the time interval of the soft proton flare observation. \textbf{The resulting X-ray spectrum is plotted in red in Figure \ref{fig:prot_SPE}: a flux of $6\times10^{-3}$ counts cm$^{-2}$ s$^{-1}$ keV$^{-1}$ is found at 1 keV, slowing decreasing towards higher energies.} The same intensity is obtained if a larger emission angle of $\pm10$ deg. is used (Figure \ref{fig:no_real}, left panel).
Only protons with energies above 50 keV are able to cross the thick filter and deposit an amount of energy below 100 keV (black spectrum of Fig. \ref{fig:prot_SPE}). If only E $<10$ keV counts are selected, the energy range of the initial proton population responsible for the observed spectrum peaks at 50 keV and rapidly drops off (Figure \ref{fig:no_real}, right panel), unlike the source of the CTE damage identified by the Chandra team that extends up to 300 keV.
\\
The reference EPIC/pn observation of a soft proton flare is the one presented by [\citenum{str00}], with the camera operated in Low Gain mode, i.e. the EPIC operation mode that decreases the gain by a $\sim10$ factor extending the sensitivity to higher energies, and the thick filter placed on. The background subtraction is performed by selecting the flaring interval and using as background the flare-free part of the observation. Only events with \textit{PATTERN}=0 are selected. The detector edges, out of field of view areas, and bright pixels are excluded from the observation. A 100\% reflection efficiency on the X-ray optics is assumed. The result is shown in blue in Fig. \ref{fig:prot_SPE}. The maximum rate, obtained below 10 keV, is $\sim1$ counts cm$^{-2}$ s$^{-1}$ keV$^{-1}$, two orders of magnitude higher than our simulated rate. Such flux is consistent with the most intense one published by Ref. \citenum{ken00}, obtained using the thin filter, but a lower intensitiy flare at $\sim0.1$ counts cm$^{-2}$ s$^{-1}$ keV$^{-1}$ is also reported.
\\
The extended soft proton induced contamination affecting the EPIC/MOS2 observation\cite{tie07} of the Lockman Hole, performed in Low Gain and with thick filter, is used to extract an X-ray spectrum of a flare. The background is subtracted using flare-free intervals as for the EPIC/pn analysis. Despite the presence of the RGS, that shades $\sim50\%$ of the incoming protons, the resulting flux is about 5 times higher than the simulated one.
Since the input Equator-S proton population is time averaged along several months, a time averaged soft proton detection should be used to compare our simulation. 
\\
\textbf{The input Equator-S proton flux of $\sim30$ prot. cm$^{-2}$ s$^{-1}$ keV$^{-1}$ sr$^{-1}$ at 50 keV translates into a simulated detection on the EPIC/pn of $6\times10^{-3}$ counts cm$^{-2}$ s$^{-1}$ keV$^{-1}$ at 1 keV: the scattering efficiency, given by the ratio of the two fluxes, is $2\times10^{-4}$ sr.}

\subsection{The impact of a magnetic diverter}\label{sec:sp_bkg}
The only solution to avoid soft proton flares for future X-ray telescopes orbiting in HEO is placing a magnetic diverter powerful enough to deflect both electrons and protons exiting the X-ray optics (see e.g., Ref. \citenum{tur06}). Missions placed in L2 could also be exposed to different but still dangereous low energy proton fluxes.
A magnetic diverter for charged particles is a device composed by a set of magnets, usually placed along the optics spider arms (not far from the exit pupil of the mirror module) to avoid vignetting, that are able to deflect, by means of the resulting tangential magnetic field, both protons and electrons, with opposite directions (see e.g., Ref. \cite{spi08}). The main issue in the development of such a device is that the proton is $\sim2000$ times heavier than the electron, so that if we take into account the curvature radius $\rho$ of a charged particle in a magnetic field:
\begin{equation}
\rho = \dfrac{1}{\rm B}\sqrt{\dfrac{2\:\rm E\:\rm m}{\rm e}}\:,
\end{equation}
where E, m and e refer to the particle energy, mass and charge respectively, protons require a magnetic field strength B$\sim40$ times higher respect with scattering electrons to achieve the same curvature radius. In [\citenum{tur06}] a deviation angle of 3.5 deg. is analytically computed to deflect protons out of a $14'$ field of view, requiring a magnetic strength B per magnetic depth l of $\sim1700$ Gauss cm.
\\
In the feasibility study of a proton diverter, the accurate knowledge of the angular, energy and spatial distribution of the protons exiting the optics is required, in order to estimate the strength of the diverter, the residual proton flux to the focal plane and the possible production of secondary background events due to the deflected proton interaction with other parts of the spacecraft. 

\section{CONCLUSIONS}
The Geant4-based BoGEMMS framework has been used to simulate the EPIC/pn X-ray background flux induced by soft proton scattering on XMM-Newton optics. The Equator-S proton flux, detected at an altitude of 70000 km in the 30-1000 keV energy range, has been used as input proton population, and the combination of Firsov (at proton incident angles $<1$ deg.) and mutiple (at proton incident angles $>1$ deg.) scattering models has been applied to model the interaction with the X-ray shells. The resulting background flux at 1 keV is $6\times10^{-3}$ counts cm$^{-2}$ s$^{-1}$ keV$^{-1}$.
\\
When comparing the simulation with the real X-ray spectrum detected by the pn and the MOS instruments of the EPIC camera in two intense soft proton flare episodes, our result underestimates up to two orders of magnitude the real case. However, it must be noted that soft proton flares are extremely variable, so that one single observation is not representive of the general behaviour. The solution would be the use of a time averaged input proton flux, and an average proton spectrum obtained by folding together soft proton flares detected along several orbits. The present results must be intended as a preliminary step in the comprehension of low energy proton scattering by X-ray optics. Simulations with the most recent Geant4 releases, providing more accurate models for the multiple scattering, and the implementation of new models for proton grazing angle scattering on the basis of laboratory measurements and ray-tracing simulations, are planned in the near future. The final aim is not only reproducing XMM-Newton soft proton flares, but more importantly the correct evaluation of this important background component for the design of future X-ray space telescopes (e.g the ESA ATHENA mission).

\bibliography{fioretti_bib} 
\bibliographystyle{spiebib} 

\end{document}